\documentclass[aps,prb,twocolumn,showpacs]{revtex4}

\usepackage{graphicx}
\usepackage{dcolumn}
\usepackage{amsmath}

\begin{document}
\bibliographystyle{apsrev}

\title{Kink nucleation in two-dimensional Frenkel-Kontorova model}
\author{Yu. N. Gornostyrev}
\author{ M. I. Katsnelson}
\author{A. V. Kravtsov }
\affiliation{Institute of Metal Physics, Ekaterinburg 620219, Russia}
\author{A. V. Trefilov}
\affiliation{Russian Science Center ''Kurchatov Institute'', Moscow 123182, Russia}

\begin{abstract}
A computer simulation of thermofluctuation nucleation of kinks on dislocations
and their dynamics is carried out in the framework of two-dimensional (2D)
Frenkel-Kontorova model. It is shown that at relatively low temperatures and
applied stresses the kinks can appear as a result of developing instability of
phonon modes localized in the vicinity of the dislocation. The transition from
this mechanism to the ordinary thermofluctuation kink nucleation with
temperature increase can reveal itself in the peculiarities of yield stress
temperature dependence.
\end{abstract}

\pacs{05.40.-a, 05.45.Yv, 61.72.Lk, 62.20.Fe}

\maketitle

\section{Introduction}

Due to the translational symmetry of crystals the energy of a dislocation
depends periodically on its position in the lattice (Peierls relief)
\cite{Hirt,Kosevitch}. This concept having been formulated more than 50 years 
ago \cite{Peierls} proved to be very fruitful for the dislocation theory, in
particular, for elucidations of factors determining the dislocation mobility 
(see, e.g., the reviews \cite{Alexander,Maeda}). At present it is commonly
accepted that at finite temperatures the dislocations in crystals glide in the
Peierls relief due to nucleation of kinks and their propagation along the
dislocation lines. This mechanism determines the dislocation mobility in covalent
crystals (like silicon) \cite{Alexander} but can also be essential in the bcc metals
and some intermetallic compounds (TiAl, NiAl) \cite{Takeuchi}.

According to the conventional point of view the velocity of dislocations 
in the Peierls relief is determined by the balance of the processes of thermofluctuation
nucleation of kinks and annihilation due to mutual recombination of kinks
and antikinks or their disappearance at the defects \cite{Alexander,Maeda}.
However, recently the experimental data have become available 
\cite{Maeda,Iunin-Nikit,Jones} which cannot be described adequately
by this simplified scheme. The results \cite{our1} of numerical simulation of
dislocation migration in the 2D Frenkel-Kontorova model also have demonstrated that 
the traditional views should be revised. As it was shown in Ref. \cite{our1} the 
kinks on the dislocations can behave like solitons and, when collising, pass through 
each other without annihilation in spite of thermal fluctuations and damping. As a 
result a much higher kink density than that predicted in terms of the traditional 
approach should be expected. It seems interesting to consider the question on the 
contribution of dynamical effects as well to the double kink nucleation since it is 
the process of kink-antikink pair nucleation that determines the temperature
dependence of the dislocation mobility \cite{Hirt}.

In the present work the process of kink nucleation is thoroughly studied by
means of modeling the dislocation movement in the 2D Frenkel-Kontorova model at
the finite temperatures. We demonstrate that, in addition to the well known mechanism of
thermoflucutation nucleation of kinks, they may also appear as a result of development
of instability of lattice vibrations localized near the dislocation axis.

\section{ Formulation of the model}

Similar to Ref. \cite{our1}, we study the dislocation dynamics in terms of the 2D
Frenkel-Kontorova (FK) model at finite temperatures. In the generalized 2D FK model
\cite{Lmdahl,our1} a layer of atoms, set into the periodic potential relief and
interacting with each other by means of elastic forces, is considered. The potential 
energy of this system has the following form
\begin{equation} 
V=\frac{K}{2}\sum_{\langle n,m\rangle}({\bf  {u}}_{n}-{\bf {u}}_{m})^{2}+
     P \sum_{n}\sum_{\bf {g}}(1-\cos{({\bf {g}{u}}_{n})})
\label{Energy}
\end{equation} 
where $K$ is the stiffness of interatomic bonds, $\langle n,m \rangle$ 
designates the sum over all the pairs of the nearest neighbors,	
${\bf g}_{1}=\frac{4\pi}{\sqrt{3}}(1; 0)$,
${\bf g}_{2,3}=\frac{4\pi}{\sqrt{3}}(\frac{1}{2} ;
\pm \frac{\sqrt{3}}{2})$ are the reciprocal lattice vectors of minimum length.

In order to consider the damping and thermal fluctuations we introduce the
contact with a thermostat using the method of Langevin  equations of motion 
\cite{vanKamp}
\begin{equation} 
\ddot{\bf  u}_{n}=-\frac{\partial{V}}{\partial{{\bf  u}_{n}}}-\gamma 
\dot{{\bf  u}}_{n}+ {\bf {\xi}}_{n}(t)+ {\bf  f}_{n}
\label{syst}
\end{equation}
where ${\bf  u}_{n}$ is the vector of displacement of the n-th atom from its 
equilibrium position, $\gamma$ is the friction coefficient, ${\bf  f}_{n}$
is the external force assumed to be equal for all the atoms, $\xi_{ni}(t)$
is the random Gaussian variable having the properties $\langle\xi_{ni}(t)\rangle=0$,  
$\langle\xi_{ni}(t)\xi_{n^{'}i^{'}}(t^{'})\rangle=2\gamma 
T\delta_{nn^{'}}\delta_{ii^{'}}\delta(t-t^{'})$, ($i=x,y$). The brackets indicate
averaging over the realization of random process  ${\bf {\xi}}_{n}(t)$. A method 
of solving the set of stochastic differential equations (\ref{syst}) is described in 
Ref. \cite{our1}.

For the numerical experiment the 2D hexagonal lattice with 40$\times$40 atoms with
periodic boundary conditions was choosen. At the initial moment the displacements in the
crystallite were specified in accordance with the known solution for dislocations in 
the 1D continual FK model \cite{Kosevitch}:
\begin{equation} 
u_{x}=0, \ \ \ 
u_{y}=1+\frac 1{\pi} \arctan\exp(-\frac{x}{\lambda}), \ \ \  
 \lambda=\frac{1}{\pi}\sqrt{\frac KP}
\label{in-cond}
\end{equation}
Then the set of equations was integrated numerically for a rather long time
($t \sim 10^2$) at ${\bf f}_{n}=0$ and the choosen temperature $T$ to find an
equilibrium configuration of the dislocation. After that an external force 
${\bf f}=(0, f_y)$, amounting to $0.75 \div 0.85$ of limiting value $f_P$ 
corresponding to the Peierls stress, was applied.

The most important parameter in the model (\ref{Energy}),(\ref{syst}) is the ratio 
of the height of potential relief $P$ to the stiffness $K$. We put $K = 1$ (which
determines the energy units) and choose $P \approx 0.1$ when, for small
displacement {\bf u}, the forces of atomic interaction with the neighbors and
with the substrate layer are comparable. This ratio of parameters leads to the 
dislocation with $\lambda \sim 1$, which is typical for most of metals 
(see \cite{Seeger,gor}). The temperature $T$ varied over a wide range from $10^{-3}$ 
to $10^{-2}$ i.e. it was of order of $(0.1 \div 0.01)P$, and the friction 
coefficient $\gamma $ varied over the range $(2\div 5)\cdot 10^{-2}$, which 
corresponds to typical values for metals at room temperatures (see, e.g., 
Ref. \cite{our1}).

\section{Computational results}

\begin{figure*}[!thb]
\begin{center}
\begin{tabular}{ccccc}
\includegraphics[width=6.50cm,clip,angle=-90]{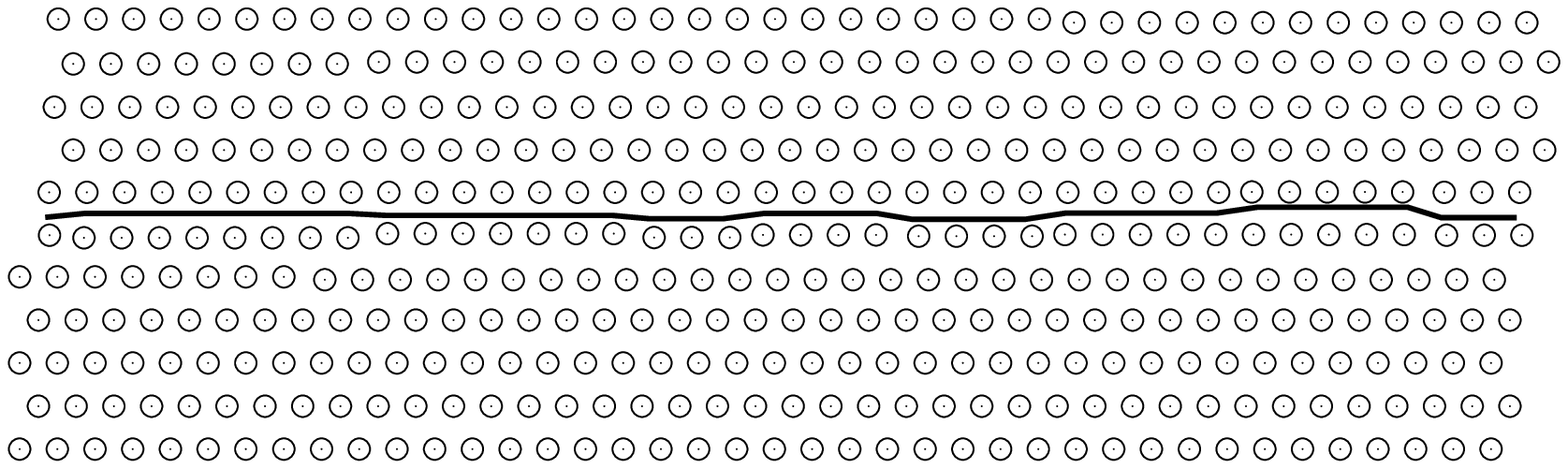}
& 
\includegraphics[width=6.50cm,clip,angle=-90]{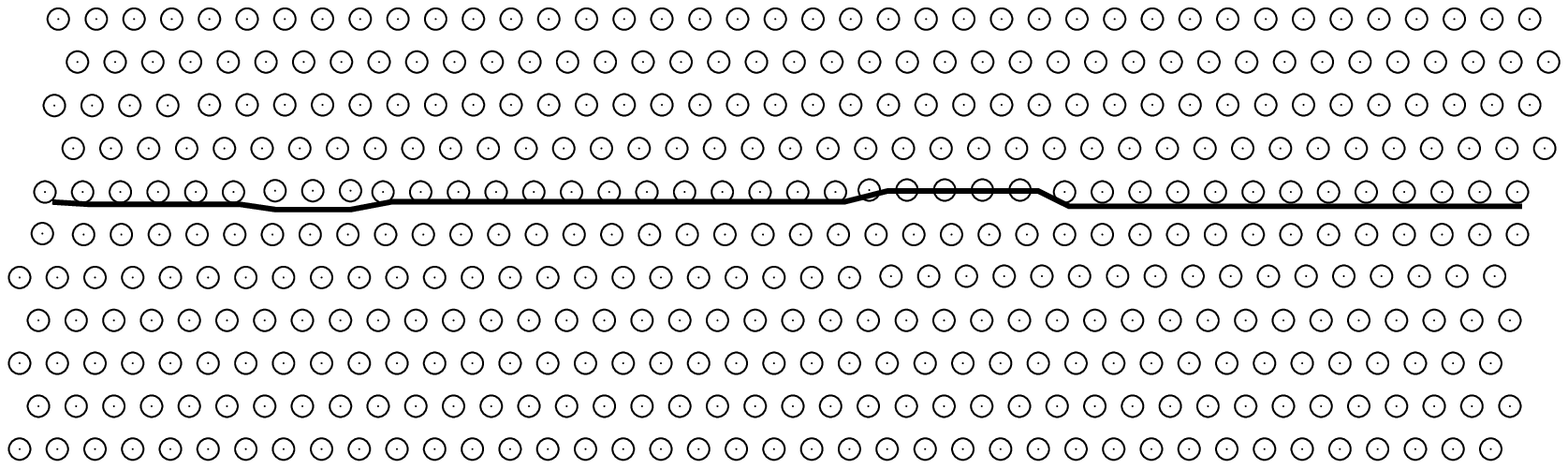}
& 
\includegraphics[width=6.50cm,clip,angle=-90]{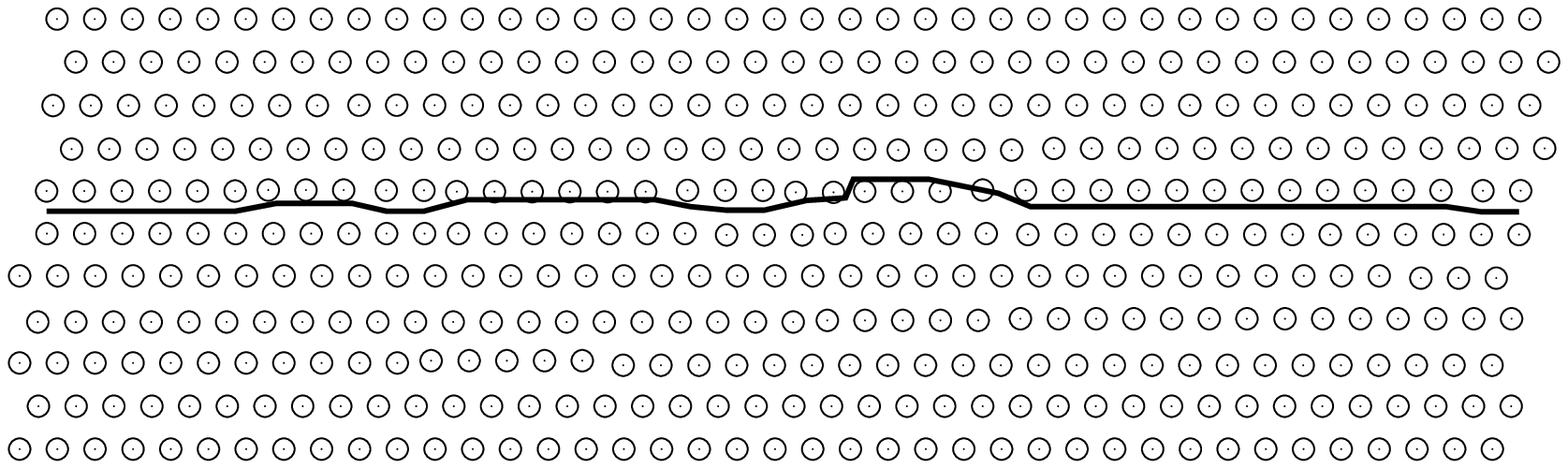}
& 
\includegraphics[width=6.50cm,clip,angle=-90]{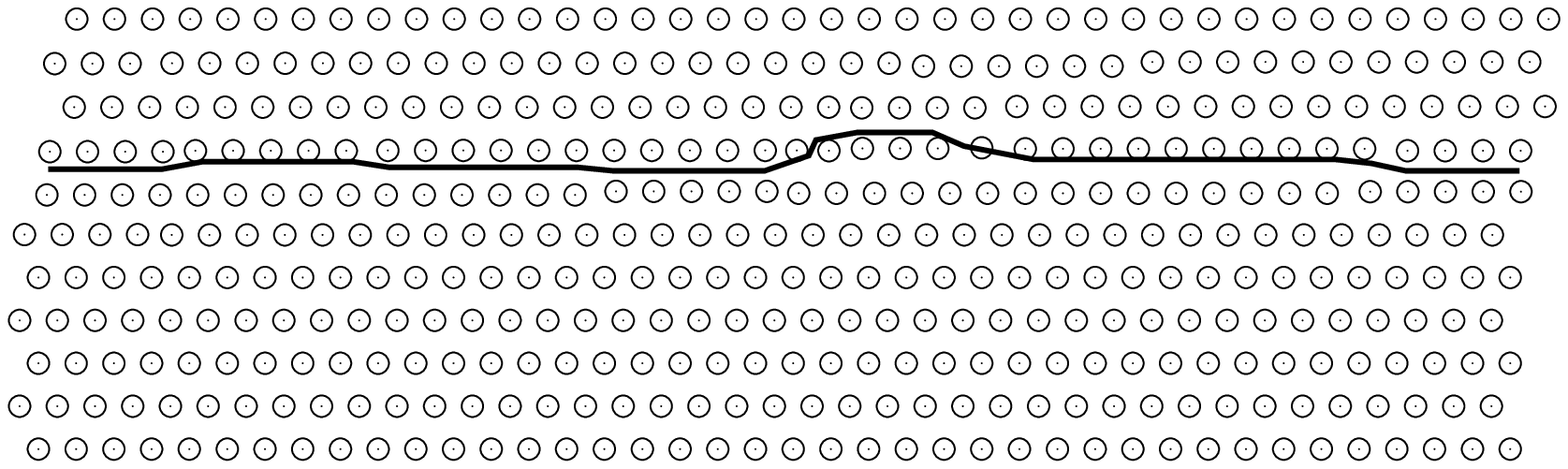}
& 
\includegraphics[width=6.50cm,clip,angle=-90]{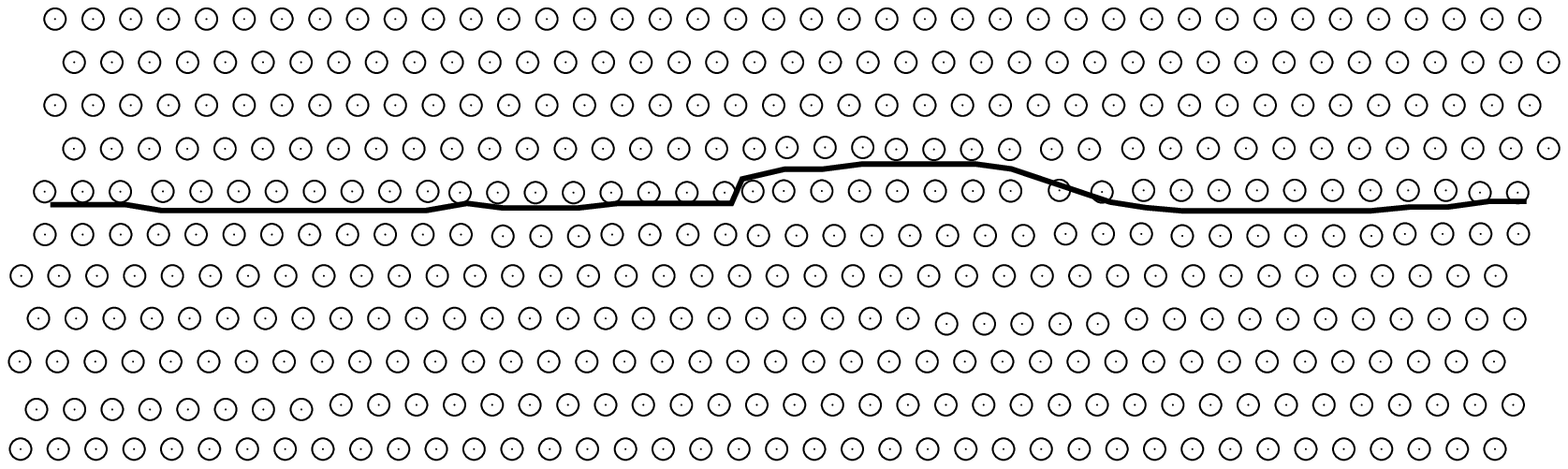} \\
$t=5$ & 13 & 14 & 15 & 17 \\
\end{tabular}
\end{center}
\vspace{-0.0cm}
\caption[a]
{
{\small
Configuration of the dislocation line (determined by the condition 
$|u_{y} | = 0.5b$) at different times for
 $P =0.1$,  $T=0.01$, $\gamma =0.02$, $f=-0.061$. 
The rows closest to the dislocation line are
$n=19,20$.
  }
}
\label{fig:Ut-hs}
\end{figure*}

Fig.\ref{fig:Ut-hs} shows a sequence of dislocation states in the 2D lattice,
demonstrating
thermofluctuation nucleation of kink-antikink pairs. It is seen that their appearance is
preceded by the stage of small amplitude oscillations of the dislocation segment. At a
certain time one of these oscillation modes begins to increase, its amplitude reaches 
the maximum value equal to the distance between the atom rows, and a kink-antikink pair
is formed. At subsequent moments of time the kinks move along the dislocation line,
reach the crystallite boundaries and, because of periodic boundary conditions, appear 
on its opposite edges. As it was demonstrated in Ref. \cite{our1}, under the conditions 
when the termofluctuation kink nucleation is possible, the kinks demonstrate a soliton-like 
character of motion and, when meeting, pass through each other without annihilation.

In order to understand the mechanism of the pair kink nucleation, let us consider
in more detail the picture of atom displacements. An important feature, accompanying
the kink nucleation at low values of external force ${\bf f}$ (or of the temperature), 
is the localization of long-wave lattice oscillations in the vicinity of the dislocation 
which manifests itself as a correlated motion of atoms in the rows nearest to the 
dislocation axis. This feature is illustrated by the dependence of the position of 
center of mass of atomic rows on time (Fig.\ref{fig:cm1}a). Curves 1,2 correspond 
to the $y$-component (parallel to the Burgers vector) of the average displacement 
$<u_y>$ for the atomic rows with $n=20$ and 19 (see Fig.\ref{fig:Ut-hs}) nearest to 
the dislocation axis, curves 3,4 for the atomic rows with $n=21$ and 18 following the 
nearest ones. One can see from Fig.\ref{fig:cm1}a that the atom row $n=20$ demonstrates 
rather regular displacement oscillations $<u_y>$ around the average value
with a period of tens of characteristic phonon times. The amplitude of these 
oscillations decreases rapidly with the distance from the dislocation axis, and for $n=18$
it is practically comparable with the thermal noise level (curve 4, Fig. \ref{fig:cm1}a). 
Note that the amplitude of the $x$-component of average displacements $<u_y>$ 
(perpendicularly to the Burgers vector) for rows $n=20$ and 19 is more than by an 
order of magnitude smaller than $<u_y>$. This reduces effectively the dimensions of the 
problem. The possibility of localization of lattice oscillations on the dislocation was 
predicted earlier \cite{Kosevitch} in terms of a simple continual model. Here this
phenomenon was found from the numerical experiment on the discrete lattice with 
allowance for the temperature and damping.

The oscillation stage ends with the nucleation of kink pair whose
propagation along the dislocation leads to monotonous growth in $<u_y>$
dependence on $t$ (curve 1, Fig.\ref{fig:str}). Thus the analysis of atom displacements 
reveals that the kink pair nucleation is due to the loss of stability of atom row motion 
in the effective periodic potential and the onset of an inhomogeneous state similar to the 
crowdion-anticrowdion pair (see Fig.\ref{fig:str}). A similar phenomenon for the 1D 
FK model was discussed in Ref. \cite{Fasolino}. It is this configuration in the row of 
atoms that corresponds to appearance of the kink-antikink pair on the dislocation line 
(Fig.\ref{fig:Ut-hs}). 
Thus, the 2D calculations demonstrate clearly that there is a region of parameters where 
the scenario of kink pair nucleation on a dislocation is due to the oscillatory 
instability near the dislocation axis rather than to the classical pattern of 
thermoflucutation formation and growth of the nucleus. In addition, the kinks on 
the dislocation in the 2D FK model are found to be similar to crowdions in the 1D 
FK model. Note that the results obtained not only support this commonly accepted 
analogy but clarify and extend it.

The observed picture of the kink pair nucleation has an uncommon non-monotonous
dependence in parameters $T, {\bf f}$ (Fig.\ref{fig:sm2}): with the rise
on temperature and/or the magnitude of applied force the probability of kink
nucleation first drastically reduces (in region II the kinks do not practically
appear during the simulation) and with further increase in the parameters are
easily formed again (region III). The diagram shown in Fig.\ref{fig:sm2} is constructed
for fixed parameters $P$ and $\gamma $; the increase in the damping results in
the reduction of the region I. The curves in Fig.\ref{fig:cm1}b illustrate
the system behavior in three different regions. Curve 1 corresponds to the kink
nucleation at low temperature and force applied, described above in detail (region I).
Curve 2 corresponds to the intermediate region II where the amplitude of oscillations
$<u_y>$ decreases rapidly with time to a value comparable with the level of thermal
noise. Curve 3 is typical of high temperatures or external forces (region III) when the
kink nucleation is again observed. However the "high-temperature" kinks differ
essentially from the "low-temperature" ones; they are much wider so that
displacements change slowly at a distance to some lattice parameters (Fig.\ref{fig:str}b).
Their nucleation is more similar to the classical pattern of fluctuation nucleation
and growth of the nucleate, without the stage of preliminary oscillations of the
dislocation line (Fig.\ref{fig:Ut-hs}) which is typical of region I.

\section{Discussion}

\begin{figure*}[!thb]
\begin{center}
\begin{tabular}{c}
a \\
\includegraphics[width=6.50cm,clip,angle=-90]{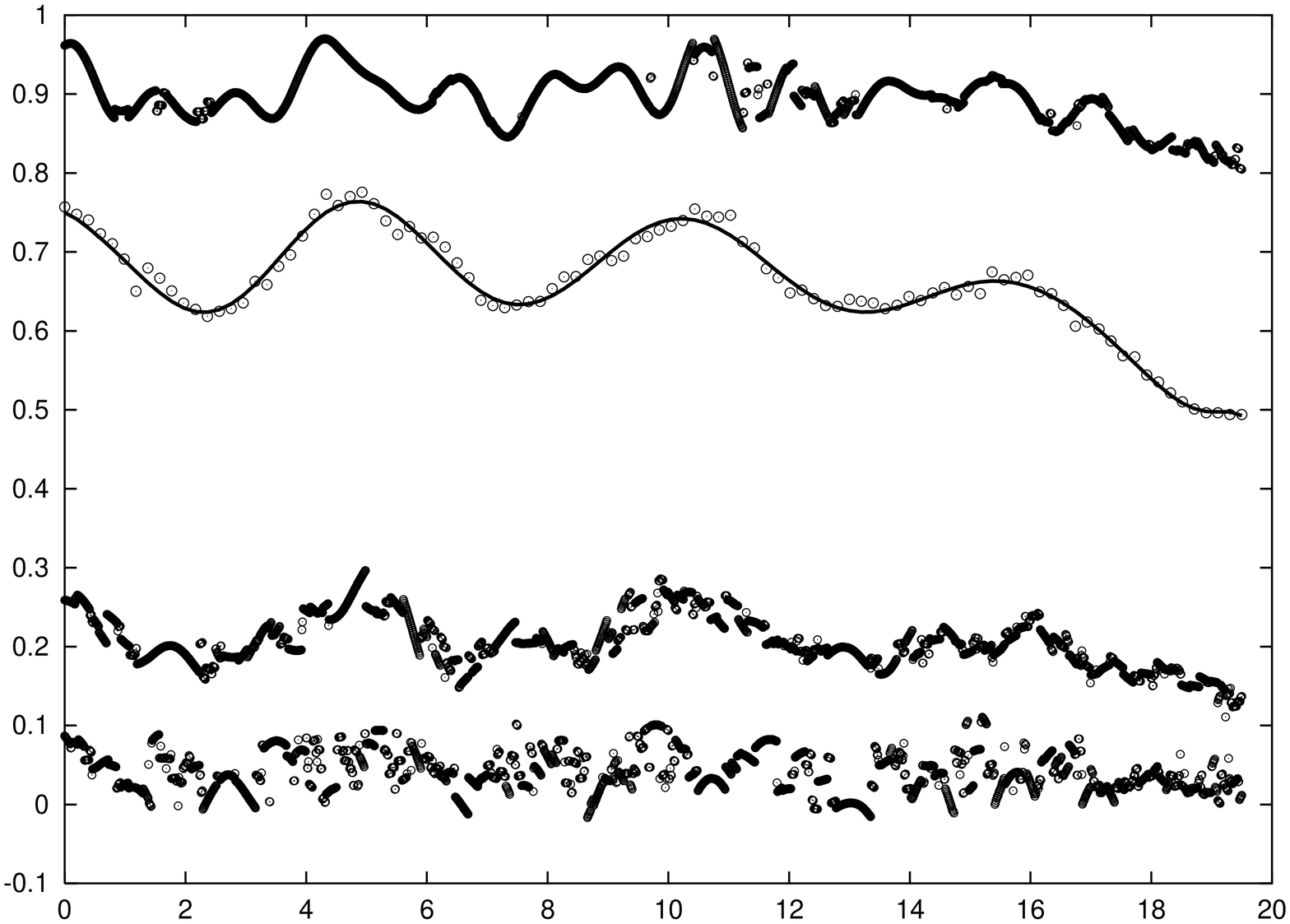} \\
b \\
\includegraphics[width=9.0cm,clip]{fig2b.eps} \\
\end{tabular}
\end{center}
\vspace{-0.0cm}
\caption[a]
{
{\small
Time dependence of the center of mass dispacement of atomic chains
for the rows nearest to the dislocation line with 
$n=20$, 21, 19, 18 (curves 1-4) at  $P=0.1, \gamma = 0.02$ $T=0.01, f=-0.061$ (a)
and for atom row with $n=20$ при $P=0.1, \gamma = 0.02$, $T=0.01, f=-0.061$
(curve 1), $T=0.08, f=-0.07$ (curve 2), $T=0.01, f=-0.07$ (curve 3) (b).
Solid line is the result of smearing of the computational data
to eliminate the thermal noise.
  }
}
\label{fig:cm1}
\end{figure*}

The computer simulations in the 2D FK model described above demonstrate a new
mechanism of kink nucleation on a dislocation as a result of development of the
instability in the phonon subsystem. Along with the phonon modes spreading within
the whole crystallite (bulk phonons) there are also modes localized near the
dislocation. In terms of continuum model the possibility of existence of such
localized oscillations was discussed earlier (see Ref. \cite{Kosevitch}). These
oscillations are clearly seen in Fig.\ref{fig:cm1}.

\begin{figure*}[!thb]
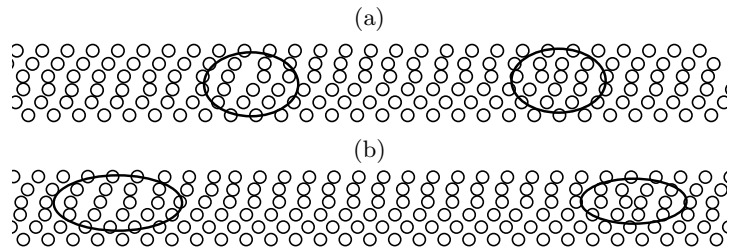

\begin{center}
\begin{tabular}{c}
(a) \\
\includegraphics[width=9.50cm,clip]{fig3a.eps}\\
(b) \\
\includegraphics[width=9.50cm,clip]{fig3b.eps}\\
\end{tabular}
\end{center}
\vspace{-0.0cm}
\caption[a]
{
{\small
Crystallite fragments showing the structure of ``low-temperature''
($T=0.003, f_y=-0.07$, a) and ``high-temperature''
($T=0.01, f_y=-0.068$, b) kinks. The regions of the compression and
expansion corresponds to crowdion and anticrowdion, correspondingly.
    }
}
\label{fig:str}
\end{figure*}

\begin{figure*}[!thb]
\centerline{
\includegraphics[width=8.50cm,clip]{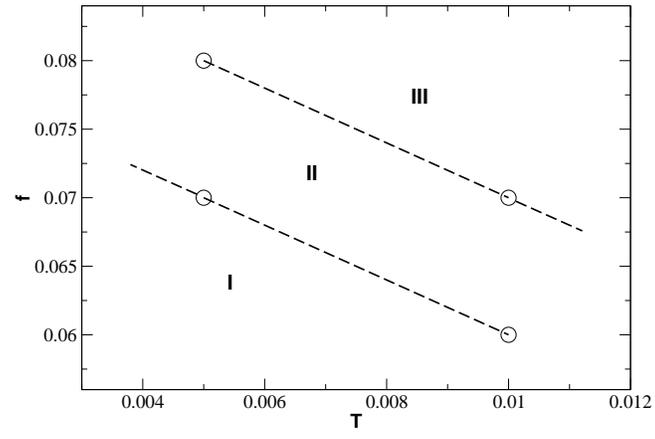}}
\vspace{-0.0cm}
\caption[a]
{
{\small
 Regions of the parameter values where the kink-antikink pair nucleation
occurs from localized oscillations (I), kinks do not nucleate (II), and they nucleate by
the standard thermofluctation mechanism (III) at $P=0.1$ и $\gamma =0.02$.
The circles show the parameter values at which the change in the mechanism was observed.
 }
}
\label{fig:sm2}
\end{figure*}

At not too high temperatures and forces (region I, Fig.\ref{fig:sm2}) the motion
of atoms in the rows nearest to the dislocation line appears to be correlated,
which manifests itself in regular oscillations $<u_y>$ (Fig.\ref{fig:cm1}a). The
change in the behavior of atom motion in the transition to the region II of
``intermediate'' temperatures and forces (curve 2, Fig.\ref{fig:cm1}) is accounted
for by the increased damping of localized phonons and loss of coherency,
first in the displacements of neighbor atom rows and then inside them. As a result the
kink nucleation turns out to be suppressed (note that the growth of damping leads to
the same consequences). With further increase in the temperature (curve 3,
Fig.\ref{fig:cm1}) the kinks again begin to nucleate but in this case it is a result
of development of the instability of ``common'' bulk phonons, which corresponds to
the traditional mechanism of thermoflucutation nucleation of the double kink
\cite{Hirt}. The transition from region I to region III is accompanied by the change
in the kink structure, from narrow kinks (Fig.\ref{fig:str}a) to the wide ones
(Fig.\ref{fig:str}b). This indicates the decrease in the secondary Peierls relief
\cite{Hirt}. The latter is that shows the loss of coherence in the atom motion in
transition to ``high-temperature'' region III.

Let us discuss the significance of the results obtained for understanding the
regularities of dislocation motion in the Peierls relief. This motion, generally
speaking, occurs due to a complicated collective process with the participation of
many kinks. We confined ourselves here to the consideration of the initial stage:
nucleation of single kink-antikink pairs. A new phenomenon of fast kink nucleation
(for the characteristics times of order of 10-100 inverse phonon frequencies) has
been found. In reality, apart from this process a slower thermoactivated kink nucleation
process takes place. In the present work we
did not study the temperature dependence of the kink nucleation rate. One can assume,
however, that for the fast processes considered here, this dependence is weaker than
for slow termoactivated processes. Therefore one can expect that the transitions
from the low temperature region to the intermediate temperature one (suppression of the
process of  kink nucleation) and from the intermediate to high temperature region
(restarting of this process) may correspond to inflection points of the temperature
dependence of yield stress which have been actually observed in some metals and alloys
with a sufficiently high Peierls relief \cite{Kuramoto,Takeuchi}. However, this question
needs further investigations.

\begin{acknowledgments}
This work is partially supported by the Netherlands Organization for Scientific 
Research, NWO project 047-008-16, by Russian Basic Research Foundation
grants 01-02-16108 and 00-15-965-44, and by Russian Science Support Foundation.
\end{acknowledgments}

\newpage

\begin{thebibliography}{99}

\bibitem{Hirt} 
 J. Hirth and J. Lothe, {\it Theory of Dislocations} (Wiley,
New York, 1982).

\bibitem{Kosevitch} 
A. M. Kosevich, In: {\it Dislocations in Solids}, ed. by.
F. R. N. Nabarro (North-Holland, Amsterdam, 1979), vol. 1, p. 33.

\bibitem{Peierls}
R. Peierls, Proc. Phys. Soc. (London) {\bf 52},  34  (1940);


\bibitem{Alexander} H. Alexander. In: {\it Dislocations in Solids}, ed. by F. R. N.
		    Nabarro (North--Holland, Amsterdam, 1986) vol. 8, p. 115.



\bibitem{Maeda}  K. Maeda and S. Takeuchi, In: {\it Dislocation in Solids}, ed. by F. R. N.
 Nabarro (North--Holland, Amsterdam, 1996), vol. 10, p. 445.


\bibitem{Takeuchi} 
S.Takeuchi, Philos. Mag. {\bf 71}, 1255 (1995).

\bibitem{Iunin-Nikit}
Yu. L. Iunin, V. I.Nikitenko, V. I. Orlov, and B. I.
Petukhov, Phys. Rev. Lett. {\bf 78}, 3137 (1997);
V. I. Nikitenko, B. Ya. Farber, and Yu. L. Iunin, Zh. Eksp. Teor. Fiz. {\bf 93}, 1304 (1987);
Yu. L. Iunin and V. I. Nikitenko, Scripta Mater. {\bf 45}, 1239 (2001).

\bibitem{Jones}
R. Jones and A. T. Blumenau, Scripta Mater. {\bf 45}, 1253 (2001)

\bibitem{our1} Yu. N. Gornostyrev, M. I. Katsnelson, A. V. Kravtsov, and
A. V. Trefilov,  Phys. Rev. {\bf B} 60, 1013 (1999).

\bibitem{Lmdahl} D. H. Srolovitz and P. S. Lomdahl, Physica D {\bf 23}, 402 (1986);
P. S. Lomdahl and D. H. Srolovitz, Phys. Rev. Lett.{\bf 57}, 2702 (1986).

\bibitem{vanKamp} 
V.G. van Kampen, {\it Stochastic Processes in Physics
and Chemistry} (North-Holland, Amsterdam, 1981).

\bibitem{Seeger} 
A. Seeger, P. Schiller 
In: {\it Physical acoustics}, ed. by U. Meson, v. 3, Part A.
(Academic press, New York, 1966)

\bibitem{gor}
Yu. N. Gornostyrev, M. I. Katsnelson, A.G. Mihin, Yu.N. Osetskii, and A. V. Trefilov,
Phys. Metals and Metallogr. {\bf 77}, 45 (1994).

\bibitem{Fasolino}
L. Consoli, H. J. F. Knops, and A. Fasolino,
Phys. Rev. Let. {\bf 85}, 302 (2000)

\bibitem{Kuramoto}
Y. Aono, K. Kitajima, E. Kuramoto, Scripta Met., {\bf 15}, 275 (1981) 


\end{thebibliography}
\end{document}